\documentclass[twocolumn,english,superscriptaddress, twocolumn, aps, prl, reprint]{revtex4-2}
\usepackage{lmodern}
\usepackage[T1]{fontenc}
\usepackage[latin9]{inputenc}
\usepackage{geometry}
\usepackage{color}
\usepackage{babel}
\usepackage{amsmath}
\usepackage{amssymb}
\usepackage{cancel}
\usepackage{graphicx}
\usepackage[pdfusetitle,
 bookmarks=false,
 breaklinks=true,pdfborder={0 0 0},pdfborderstyle={},backref=false,colorlinks=true]
 {hyperref}
\hypersetup{
 pdfstartview=XYZ}

\makeatletter
\@ifundefined{date}{}{\date{}}

\usepackage{xcolor}
\usepackage{wasysym}
\usepackage{tikz}
\usepackage{esint}
\def\tagform@#1{\maketag@@@{\selectlanguage{english}(\ignorespaces#1\unskip\@@italiccorr)}}
\usepackage{cleveref}

\definecolor{crimson}{RGB}{220,20,60}
\hypersetup{citecolor=magenta,
    urlcolor=cyan}
\usepackage{csquotes}

\renewcommand\[{\begin{equation}}
\renewcommand\]{\end{equation}}

\AtBeginDocument{
	\renewcommand{\hbar}{{h\hspace*{-0.3em}\bar{}\hspace*{0.268em}}}
}

\usepackage[caption=false]{subfig}

\ifdefined\showcaptionsetup
 \PassOptionsToPackage{caption=false}{subfig}
\fi
\usepackage{subfig}
\AtBeginDocument{
  
}

\makeatother

\begin{document}
\title{Unvortex Lattice and Topological Defects in Rigidly Rotating Multicomponent
Superfluids}
\author{Roy Rabaglia}
\affiliation{Department of Physics, Technion, Haifa 32000, Israel}
\author{Ryan L. Barnett}
\affiliation{Department of Mathematics, Imperial College London, London SW7 2AZ,
United Kingdom}
\author{Ari M. Turner}
\affiliation{Department of Physics, Technion, Haifa 32000, Israel}
\date{\today}
\begin{abstract}
\noindent By examining rotating ferromagnetic spinor condensates through
the perspective of large spin, we identify a novel type of topological
point defects in the magnetization texture. These defects are not
predicted by conventional homotopy analysis but rather by the Riemann--Hurwitz
formula. The magnetization texture in the system is described by an
equal-area mapping from the plane to the sphere of magnetization,
forming a lattice of uniformly charged Skyrmions. This lattice contains
doubly-quantized (winding number $=2$) point defects arranged on
the sphere in a tetrahedral configuration. The fluid found to be
rotating rigidly, except at the point defects, where the vorticity
vanishes. This vorticity structure describes an unconventional ``unvortex''
lattice, which contrasts with the well-known vortex lattice in scalar
rotating superfluids, where vorticity is concentrated exclusively
within defect points. Numerical results are presented, confirming
these predictions and demonstrating their persistence in smaller-spin
condensates.
\end{abstract}
\maketitle
One fundamental characteristic of ordinary (scalar) superfluids is
their irrotational flow. In contrast, spinor Bose--Einstein condensates
exhibit an intrinsic coupling between the superfluid velocity field
$\mathbf{v}\left(\mathbf{r}\right)$ and the magnetization unit vector
field $\hat{\mathbf{n}}\left(\mathbf{r}\right)$. This coupling is
governed by the well-known \emph{Mermin-Ho relation} \citep{Circulation=000020and=000020angular=000020momentum=000020in=000020the=000020a=000020phase=000020of=000020superfluid=000020Helium-3}:
\begin{equation}
\left(\boldsymbol{\nabla}\times\mathbf{v}\right)_{k}=\frac{\hbar}{2m}F\varepsilon_{ijk}\hat{\mathbf{n}}\cdot\left(\partial_{i}\hat{\mathbf{n}}\times\partial_{j}\hat{\mathbf{n}}\right),\label{eq:=000020MH}
\end{equation}
where $m$ and $F$ are the mass and spin of the condensed particles,
respectively. For a planar ($d=2$) condensate, this relation has
an intriguing geometrical interpretation: the right-hand side is proportional
to the Jacobian $J\left(\mathbf{r}\right)$ of the transformation
$\hat{\mathbf{n}}\left(\mathbf{r}\right):\mathbb{R}^{2}\rightarrow\mathbb{S}^{2}$,
mapping the physical space to the sphere of spin states.

A rich variety of magnetization textures and flow fields are known
to occur in spinor condensates \citep{Spinor=000020Bose=002013Einstein=000020condensates,Two-component=000020Bose-Einstein=000020condensates=000020with=000020a=000020large=000020number=000020of=000020vortices,Rotating=000020spin-1=000020bosons=000020in=000020the=000020lowest=000020Landau=000020level,Einstein=002013de=000020Haas=000020effect=000020in=000020dipolar=000020Bose-Einstein=000020condensates,Spontaneous=000020spin=000020textures=000020in=000020dipolar=000020spinor=000020condensates,Solitons=000020in=000020a=000020trapped=000020spin-1=000020atomic=000020condensate,Rotating=000020dipolar=000020spin-1=000020Bose=002013Einstein=000020condensates,Neutral=000020skyrmion=000020configurations=000020in=000020the=000020low-energy=000020effective=000020theory=000020of=000020spinor-condensate=000020ferromagnets,Ground=000020states=000020solitons=000020and=000020spin=000020textures=000020in=000020spin-1=000020Bose-Einstein=000020condensates,Route=000020to=000020non-Abelian=000020quantum=000020turbulence=000020in=000020spinor=000020Bose-Einstein=000020condensates,Dynamics=000020and=000020complex=000020structure=000020of=000020two-dimensional=000020skyrmions=000020in=000020antiferromagnetic=000020spin-1=000020bose-einstein=000020condensates,Stable=000020core=000020symmetries=000020and=000020confined=000020textures=000020for=000020a=000020vortex=000020line=000020in=000020a=000020spinor=000020bose-einstein=000020condensate,Two-dimensional=000020solitons=000020and=000020quantum=000020droplets=000020supported=000020by=000020competing=000020self-and=000020cross-interactions=000020in=000020spin-orbit-coupled=000020condensates,Spin=000020and=000020mass=000020superfluidity=000020in=000020a=000020ferromagnetic=000020spin-1=000020Bose-Einstein=000020condensate,Skyrmions=000020with=000020arbitrary=000020topological=000020charges=000020in=000020spinor=000020Bose=002013Einstein=000020condensates,Three-Dimensional=000020Skyrmions=000020with=000020Arbitrary=000020topological=000020number=000020in=000020a=000020Ferromagnetic=000020Spin-1=000020Bose-einstein=000020condensate,Existence=000020stability=000020and=000020dynamics=000020of=000020monopole=000020and=000020Alice=000020ring=000020solutions=000020in=000020antiferromagnetic=000020spinor=000020condensates,Half-quantum=000020vortices=000020in=000020an=000020antiferromagnetic=000020spinor=000020Bose-Einstein=000020condensate,Dynamics=000020of=000020a=000020vortex=000020dipole=000020across=000020a=000020magnetic=000020phase=000020boundary=000020in=000020a=000020spinor=000020Bose-Einstein=000020condensate}.
One such phenomenon, facilitated by the Mermin-Ho relation, is the
elimination of the need for vortices with diverging velocity \citep{Coreless=000020vortex=000020ground=000020state=000020of=000020the=000020rotating=000020spinor=000020condensate,Coreless=000020and=000020singular=000020vortex=000020lattices=000020in=000020rotating=000020spinor=000020Bose-Einstein=000020condensates,Coreless=000020vortex=000020formation=000020in=000020a=000020spinor=000020Bose-Einstein=000020condensate}.
In its ground state, a non-rotating ferromagnetic condensate features
uniform magnetization. Upon rotation, maintaining uniform magnetization
would lead to an irrotational flow around a lattice of quantized vortices,
as in ordinary superfluids. However, the Mermin-Ho relation enables
the system to reduce its energy by adopting non-uniform magnetization,
leading to a non-trivial flow field \citep{Two-component=000020Bose-Einstein=000020condensates=000020with=000020a=000020large=000020number=000020of=000020vortices,Rotating=000020spin-1=000020bosons=000020in=000020the=000020lowest=000020Landau=000020level,Spin-orbit=000020coupled=000020Bose-Einstein=000020condensate=000020under=000020rotation,Energetically=000020stable=000020singular=000020vortex=000020cores=000020in=000020an=000020atomic=000020spin-1=000020Bose-Einstein=000020condensate,Vortices=000020in=000020multicomponent=000020Bose=002013Einstein=000020condensates}.
While a uniform circulation matching rigid rotation might seem energetically
favorable, we show that a new class of defects induces large variations
in circulation.

This discussion is based on a novel analytical approach to understand
properties of spinor condensates, by inspecting the system from the
viewpoint of large spin ($F\gg1$). Experimental realizations of spinor
condensates have so far reached total spin values up to $F=8$ \citep{All-optical=000020formation=000020of=000020an=000020atomic=000020Bose-Einstein=000020condensate,Observation=000020of=000020spinor=000020dynamics=000020in=000020optically=000020trapped=000020rb=00002087=000020bose-einstein=000020condensates,Optical=000020confinement=000020of=000020a=000020Bose-Einstein=000020condensate,Observation=000020of=000020metastable=000020states=000020in=000020spinor=000020Bose-Einstein=000020condensates,Spinor=000020dynamics=000020in=000020an=000020antiferromagnetic=000020spin-1=000020condensate,Observation=000020of=000020a=000020strongly=000020ferromagnetic=000020spinor=000020Bose-Einstein=000020condensate,Magnetic=000020field=000020dependence=000020of=000020the=000020dynamics=000020of=000020Rb=00002087=000020spin-2=000020Bose-Einstein=000020condensates,Dynamics=000020of=000020F=00003D=0000202=000020spinor=000020Bose-Einstein=000020condensates,Sodium=000020Bose-Einstein=000020condensates=000020in=000020the=000020F=00003D=0000202=000020state=000020in=000020a=000020large-volume=000020optical=000020trap,All-optical=000020production=000020of=000020chromium=000020Bose-Einstein=000020condensates,Spontaneous=000020demagnetization=000020of=000020a=000020dipolar=000020spinor=000020Bose=000020gas=000020in=000020an=000020ultralow=000020magnetic=000020field,Bose-Einstein=000020condensation=000020of=000020erbium,Bose-Einstein=000020condensation=000020of=000020europium,Strongly=000020dipolar=000020Bose-Einstein=000020condensate=000020of=000020dysprosium}.
This range already allows exploration of the large-$F$ behavior discussed
here. We show that the large spin viewpoint is capable not only of
capturing the properties of the larger spin condensates, but can even
provide a framework for understanding smaller-spin condensates, such
as spin $1$. Expansions around $F=\infty$ can be combined with results
of expansions around $F=0$, which we will present in future work
\citep{future}, to provide a fairly accurate description of condensates
of any spin value.

\begin{figure*}[t]
\centering
\begin{centering}
\subfloat[]{\begin{centering}
\includegraphics[scale=0.21]{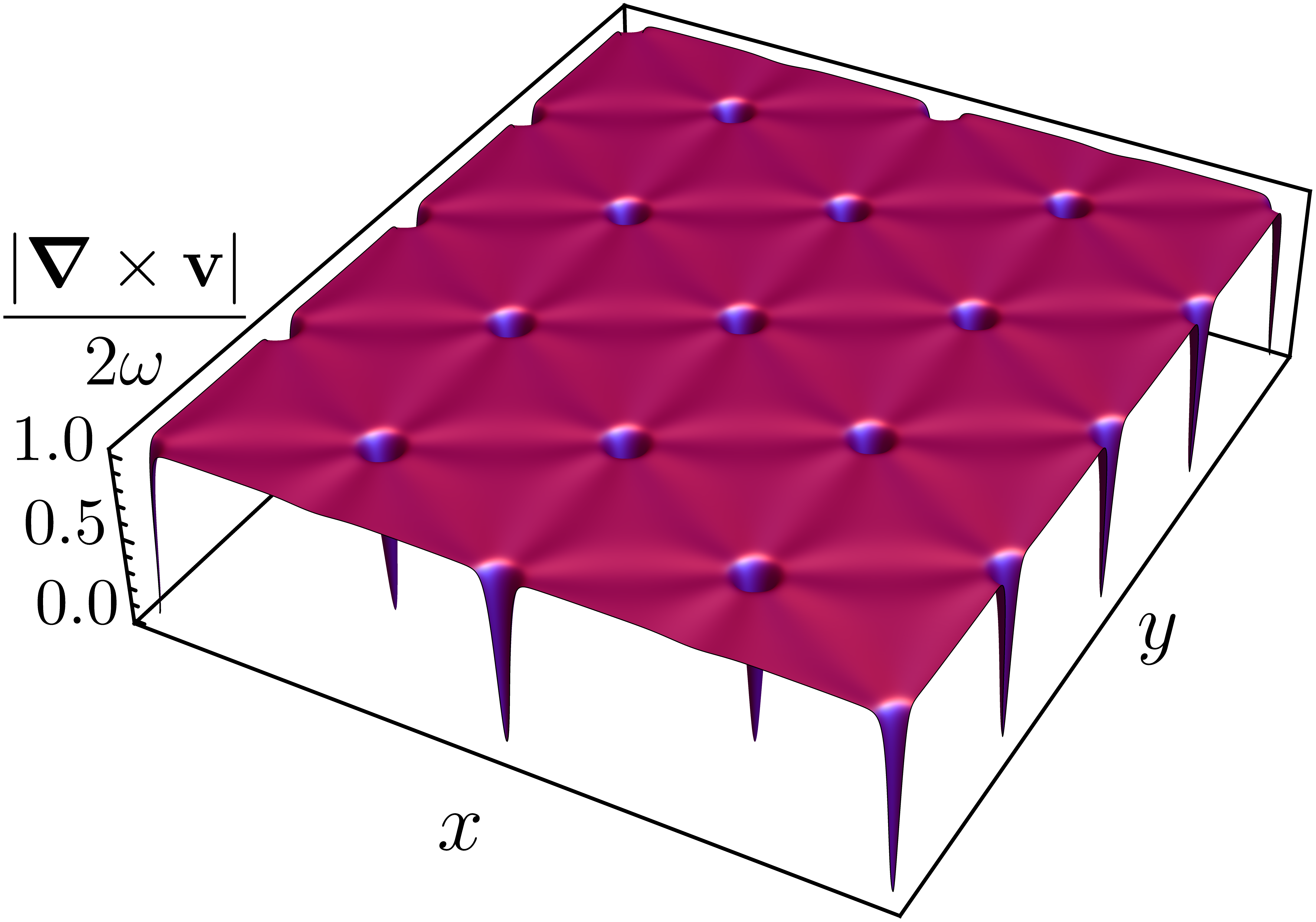}
\par\end{centering}
}\subfloat[]{\begin{centering}
\includegraphics[scale=0.21]{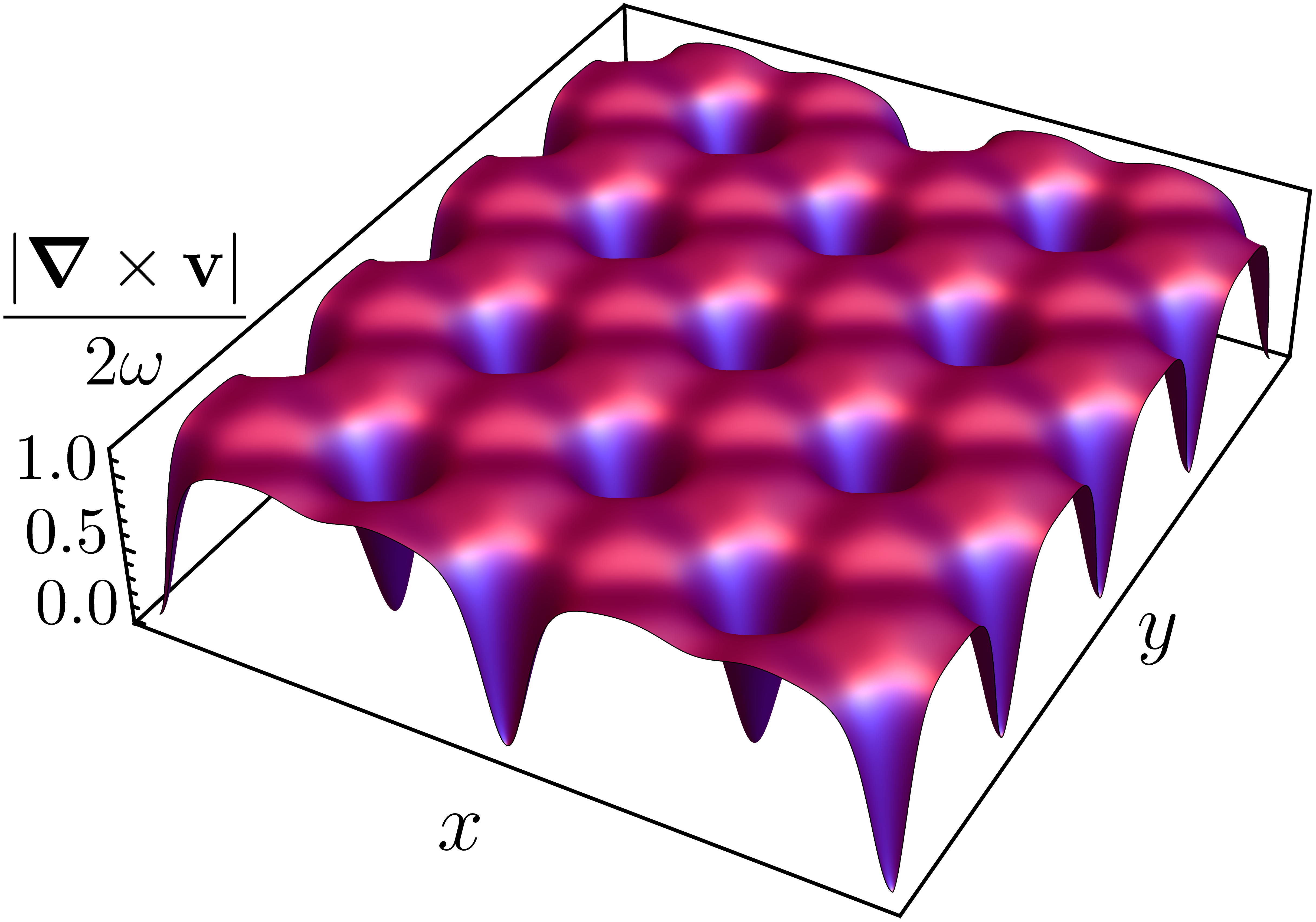}
\par\end{centering}
}\subfloat[]{\begin{centering}
\includegraphics[scale=0.21]{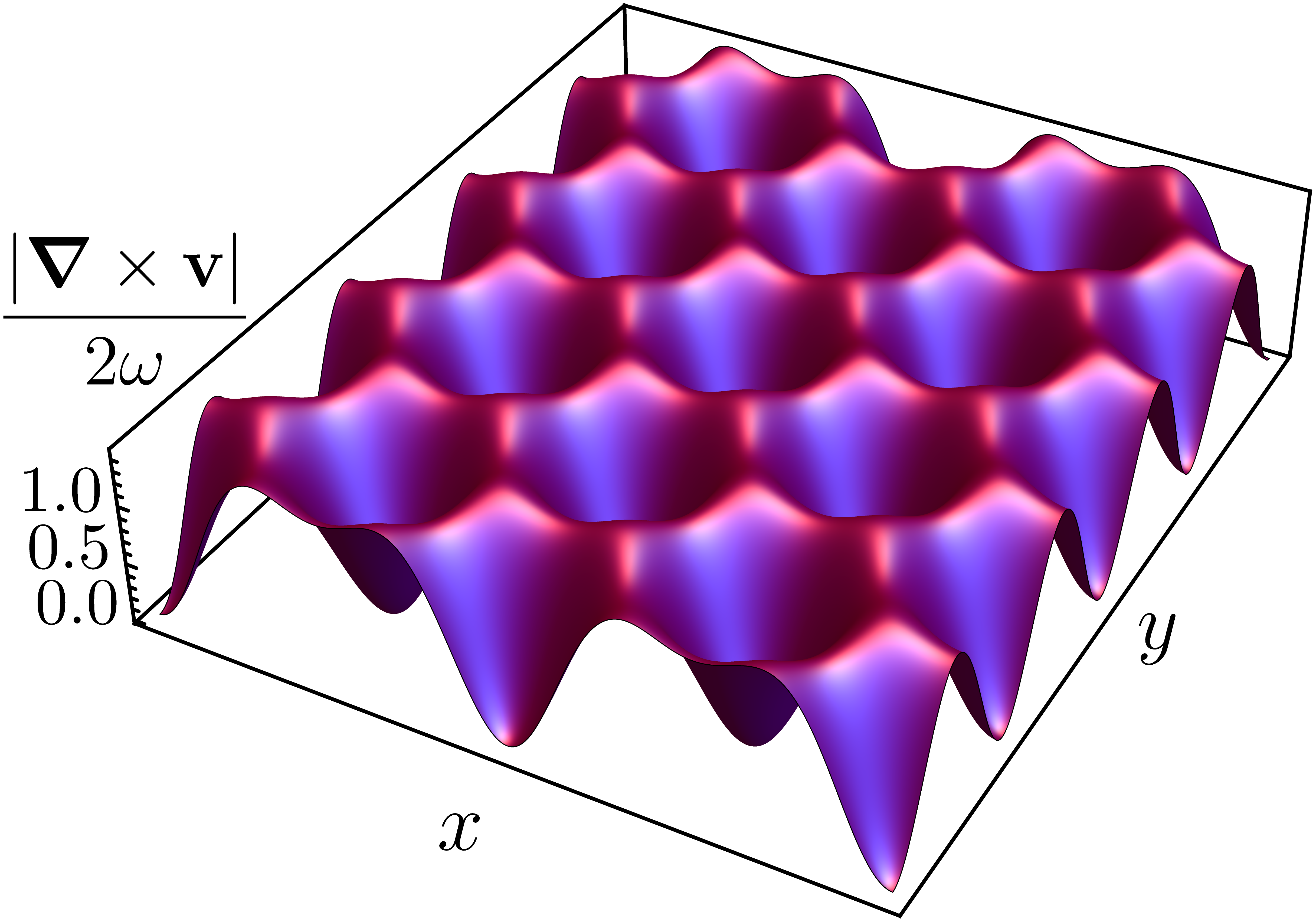}
\par\end{centering}
}
\par\end{centering}
\caption{The \textquotedblleft unvortex\textquotedblright{} lattice, obtained
numerically for (a): $F=100$, (b): $F=8$, (c): $F=1$. A triangular
lattice of cores forms around each point defect, which are predicted
by the Riemann-Hurwitz formula. For large $F$, the condensate rotates
rigidly between the cores.\protect\label{fig:The--lattice.}}
\end{figure*}

We use this system as a framework to introduce a unique kind of topological
defects, which constitute the central focus of this paper. These defects,
which emerge in the magnetic texture, are naturally understood from
the large-$F$ viewpoint, yet they persist in systems regardless of
the spin value. Their impact on the system is profound, influencing
various properties beyond the magnetic texture itself, such as the
aforementioned regions of depleted vorticity, which form around the
magnetic defects and disrupt an otherwise rigid flow. The resulting
vorticity structure (see Fig. \eqref{fig:The--lattice.}), stands
as an antithesis to the traditional vortex lattice in superfluids,
where all the vorticity is concentrated within the defects.

These defects arise from a mechanism distinct from that of standard
topological defects and do not fit within the standard classification
schemes such as homotopy theory; they result from the Riemann-Hurwitz
formula, a relationship that constrains mappings between topologically
inequivalent spaces. Similar to the planar case, these defects also
persist in the ground state of three-dimensional condensates, forming
a lattice of line defects. These can be studied for behaviors such
as bending or knotting, akin to defects in conventional superfluids
\citep{Creation=000020and=000020dynamics=000020of=000020knotted=000020vortices,Vortex=000020knots=000020in=000020a=000020Bose-Einstein=000020condensate,How=000020superfluid=000020vortex=000020knots=000020untie}.
We believe that this special type of defects might also be relevant
to physical systems far beyond spinor condensates, such as quantum
Hall systems, where skyrmions emerge and their number is fixed by
the deviation from special filling fractions \citep{Skyrmion=000020lattice=000020melting=000020in=000020the=000020quantum=000020Hall=000020system},
or blue phase of liquid crystals in which the energy is minimized
by having a texture of the nematic that varies in a non-coplanar way,
where results analogous to the Mermin-Ho relation can be used \citep{Crystalline=000020liquids:=000020the=000020blue=000020phases,Predicting=000020a=000020polar=000020analog=000020of=000020chiral=000020blue=000020phases=000020in=000020liquid=000020crystals}.

Spinor condensates can exhibit various phases determined by the interatomic
interaction parameters \citep{Spinor=000020Bose=002013Einstein=000020condensates,Low-energy=000020dynamics=000020of=000020spinor=000020condensates,Topological=000020interface=000020physics=000020of=000020defects=000020and=000020textures=000020in=000020spinor=000020Bose-Einstein=000020condensates}.
Our focus is on the ferromagnetic phase, which can occur for any spin
$F$ within a certain range of the spin-dependent interaction parameters.
Assuming a constant density profile \citep{Long-wavelength=000020spin=000020dynamics=000020of=000020ferromagnetic=000020condensates},
the condensate in this phase can be described solely by the two fields
$\mathbf{v}\left(\mathbf{r}\right)$ and $\hat{\mathbf{n}}\left(\mathbf{r}\right)$.
An alternative description involves angles: the superfluid phase $\theta\left(\mathbf{r}\right)$,
and the magnetic polar and azimuthal angles $\phi\left(\mathbf{r}\right)$
and $\chi$$\left(\mathbf{r}\right)$. Using these angles, the magnetization
texture is $\hat{\mathbf{n}}=\left(\sin\phi\cos\chi,\sin\phi\sin\chi,\cos\phi\right)$,
and the velocity field in the non-rotating frame of reference is
\begin{equation}
\mathbf{v}\left(\mathbf{r}\right)=\frac{\hbar}{m}\left[\boldsymbol{\nabla}\theta-F\cos\phi\boldsymbol{\nabla}\chi\right]\label{eq:velF}
\end{equation}
\citep{Persistent=000020currents=000020in=000020ferromagnetic=000020condensates}.
Taking the curl yields the Mermin-Ho relation, Eq. \eqref{eq:=000020MH}:
\begin{equation}
\boldsymbol{\nabla}\times\mathbf{v}=\frac{\hbar}{m}FJ\left(\mathbf{r}\right)\hat{\mathbf{z}}.\label{eq:mermin}
\end{equation}
In terms of $\phi$ and $\chi$, the Jacobian can be written as $J\left(\mathbf{r}\right)=\sin\phi\left(\boldsymbol{\nabla}\phi\times\boldsymbol{\nabla}\chi\right)\cdot\hat{\mathbf{z}}$.
When rotated with an angular velocity of $\mathbf{\boldsymbol{\omega}}=\omega\hat{\mathbf{z}}$,
the energy functional of the system is
\begin{equation}
E=\frac{\hbar^{2}\rho}{2m}\int\mathrm{d}^{2}r\bigg[\frac{m^{2}}{\hbar^{2}}\big(\mathbf{v}-\mathbf{\boldsymbol{\omega}}\times\mathbf{r}\big)^{2}+\frac{1}{2}F\big(\boldsymbol{\nabla}\hat{\mathbf{n}}\big)^{2}\bigg]\label{eq:eneF}
\end{equation}
\citep{Long-wavelength=000020spin=000020dynamics=000020of=000020ferromagnetic=000020condensates,Persistent=000020currents=000020in=000020ferromagnetic=000020condensates,foot-1}.
We want to identify the important contributions in the large-$F$
limit. Although a cursory examination of the energy functional may
seem to suggest a dominance of the second (magnetic) term, this overlooks
the $F$ dependence of the velocity field, as indicated in Eq. \eqref{eq:velF}.
This expression for the velocity field contains a term proportional
to $F$, hence the first (kinetic) term in the energy contains terms
proportional to $F^{2}$. This surprisingly indicates its dominance
over the magnetic term, which is proportional only to $F$. To show
this rigorously, we propose employing a rescaling technique.

We rescale the lengths using $\tilde{\mathbf{r}}=F^{-1/2}\mathbf{r}$
to eliminate the dependence of the magnetic term on $F$, and the
phase using $\tilde{\theta}=F^{-1}\theta$ to simplify the resulting
expression. After these rescalings, the velocity field becomes
\begin{equation}
\tilde{\mathbf{v}}=\frac{1}{\sqrt{F}}\mathbf{v}=\frac{\hbar}{m}\big[\tilde{\boldsymbol{\nabla}}\tilde{\theta}-\cos\phi\tilde{\boldsymbol{\nabla}}\chi\big],\label{eq:velocity}
\end{equation}
and contains no explicit dependence on $F$. As velocity measures
the change in position over time, it was also rescaled to account
for the rescaling of lengths. The energy, which is similarly rescaled
by $\tilde{E}=F^{-1}E$, becomes
\begin{equation}
\tilde{E}=\frac{\hbar^{2}\rho}{2m}\int\mathrm{d}^{2}\tilde{r}\bigg[\frac{m^{2}}{\hbar^{2}}F\big(\tilde{\mathbf{v}}-\mathbf{\boldsymbol{\omega}}\times\tilde{\mathbf{r}}\big)^{2}+\frac{1}{2}\big(\tilde{\boldsymbol{\nabla}}\hat{\mathbf{n}}\big)^{2}\bigg].\label{eq:energy}
\end{equation}
As we will use only the rescaled quantities, we omit the tilde symbol
in subsequent equations. Unlike the original energy functional \eqref{eq:eneF},
the rescaled energy functional exhibits a remarkably simple dependence
on $F$. After rescaling, it is clear that the dominant term in the
large-$F$ limit is the kinetic one.  Deviations from a state minimizing
this term result in a high energetic cost, implying that the condensate
rotates rigidly, such that $\mathbf{v}=\mathbf{\boldsymbol{\omega}}\times\mathbf{r}$.
A previous proposal to realize rigidly-rotating superfluids involves
the use of spin-orbit coupling \citep{Diffused=000020vorticity=000020and=000020moment=000020of=000020inertia=000020of=000020a=000020spin-orbit=000020coupled=000020Bose-Einstein=000020condensate}.
Our study predicts a natural occurrence of this phenomenon in large-$F$
spinor condensates.

The vorticity for rigid rotation is a constant, $2\mathbf{\boldsymbol{\omega}}$,
implying a constant Jacobian $J\left(\mathbf{r}\right)=2m\omega/\hbar F$
according to the Mermin-Ho relation \eqref{eq:mermin}. Therefore,
the mapping $\hat{\mathbf{n}}\left(\mathbf{r}\right)$ must be \emph{area-preserving}
(up to a constant scaling factor). We will assume $\omega>0$, and
therefore $J\left(\mathbf{r}\right)$ is positive.

Such a mapping can also be considered as describing a system with
uniform Skyrmionic charge density, where the Skyrmionic charge $Q=1/4\pi\int J\left(\mathbf{r}\right)\mathrm{d}^{2}r$
is the number of times the sphere is covered by the mapping in each
unit cell \citep{Vortices=000020in=000020multicomponent=000020Bose=002013Einstein=000020condensates,Solitons=000020and=000020instantons,Skyrmions=000020in=000020condensed=000020matter}.
This provides us with an equation for the area $A$ of the unit cell
in terms of $Q$:
\begin{equation}
A=\frac{2\pi\hbar}{m\omega}FQ.\label{eq:intMH}
\end{equation}

Any periodic, area-preserving mapping from the plane to the sphere,
or more generally, any mapping with a single-sign Jacobian, must have
defects.  These defects can appear in various forms, such as lines
or points. Our focus is on point defects, as other types of singularities
incur a large energetic cost. The defects of the type we consider
will occur in a mapping from a two-dimensional system to a two-dimensional
order parameter space (but can be generalized to higher dimensions).
We assume that a mapping $\hat{\mathbf{n}}:P\rightarrow S$ with a
positive Jacobian is favored. For each point like this, a sufficiently
small region around it will be mapped in a one-to-one way to a region
of the order parameter space. Conversely, a defect is a point $p$
with the property that a small disk around $p$, with $p$ removed
from it, is mapped in a $k$-to-one way to the order parameter space,
for $k\geq2$. As a result, the Jacobian must vanish in these specific
points, as the mapping is not invertible in their neighborhood. These
defects have the same topological structure as branching points of
analytic mappings.

\begin{figure}[h]
\centering
\noindent\includegraphics[scale=0.7]{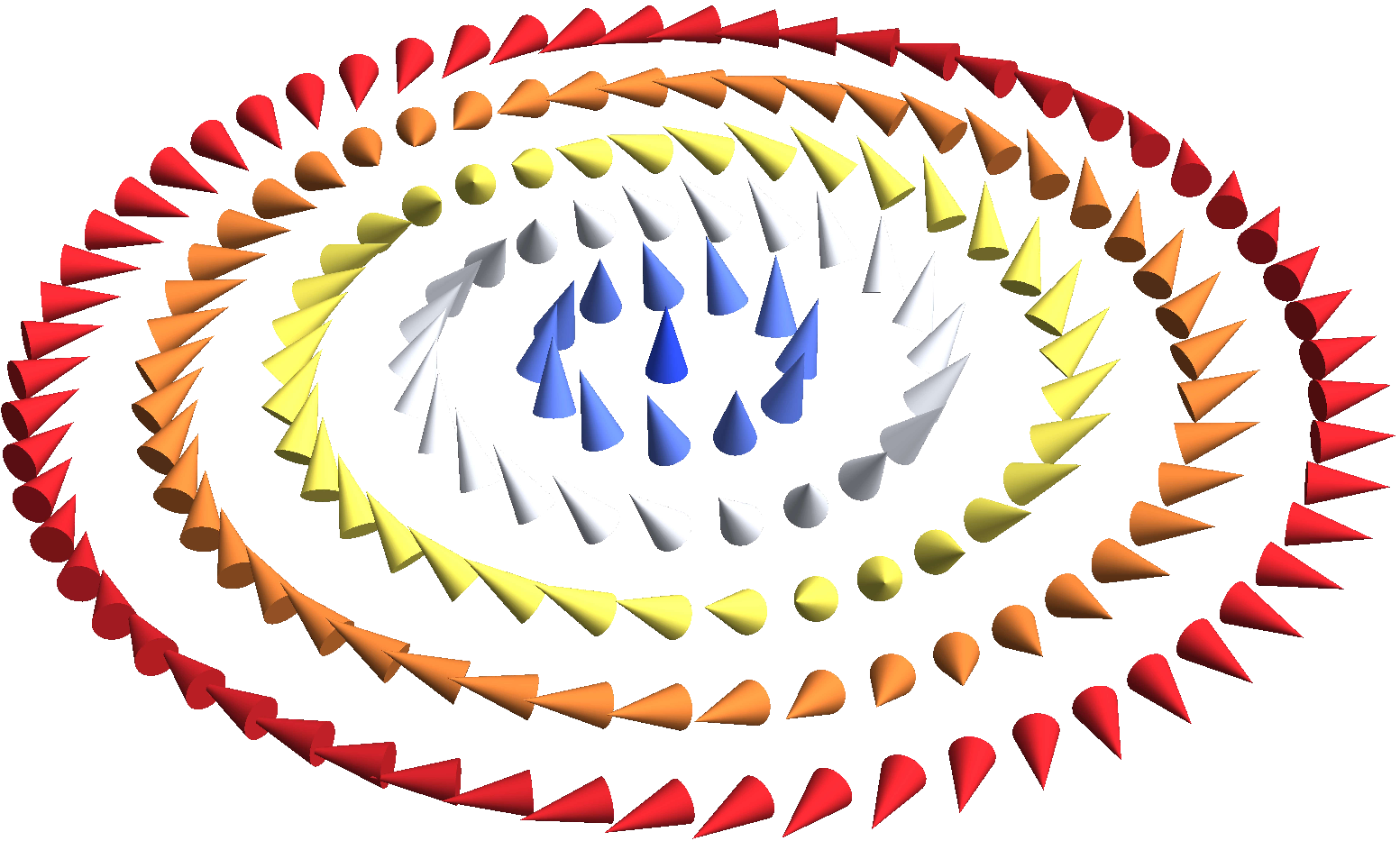}

\caption{An illustration of the magnetic texture \eqref{eq:texture=000020exmpl}
for $k=2$. Each magnetization vector is the image of two different
points, except for the magnetization of the defect itself at the origin.\protect\label{fig:An-illustration-for}}
\end{figure}

An example of such a defect is at the origin of the following texture,
using the polar coordinates $\left(r,\alpha\right)$:
\begin{equation}
\hat{\mathbf{n}}\left(\mathbf{r}\right)\hspace*{-1bp}=\hspace*{-1bp}\left(\sin\phi\left(r\right)\cos\left(k\alpha\right),\sin\phi\left(r\right)\sin\left(k\alpha\right),\cos\phi\left(r\right)\right)\label{eq:texture=000020exmpl}
\end{equation}
where $\phi\left(r\right)\rightarrow0$ as $r\rightarrow0$. This
field corresponds to the azimuthal angle $\chi=k\alpha$, having a
winding number of $k$ for this angle. As can be seen in Fig. \eqref{fig:An-illustration-for},
the winding of $\chi$ does not imply a discontinuity in the texture
$\hat{\mathbf{n}}\left(\mathbf{r}\right)$, unlike in an ordinary
vortex, since the magnetization points toward the north pole near
the origin. Yet, any set of $k$ points that are angularly spaced
by $\Delta\alpha=2\pi/k$ for the same $r$ all have the same magnetization
$\hat{\mathbf{n}}$ at them, showing that this is a $k$-to-$1$ texture,
which constitutes a $k$-defect at the origin. Although this field
is continuous, it is not possible to smooth the defect by altering
the orientation of the vectors while keeping a single-signed Jacobian;
the defect will merely be relocated to another position. 

The presence of these defects in the system is dictated by the Riemann-Hurwitz
formula, a topological theorem that establishes a relationship between
the branching points in a mapping and the topological properties of
the spaces it connects. We will assume that, away from defects, the
mapping is differentiable and has a positive Jacobian. Let $P$ and
$S$ be two closed Riemann surfaces, and let $\hat{\mathbf{n}}:P\rightarrow S$
represent the mapping between the spaces. Then each point of $S$,
except for images of defects, must be mapped to the same number of
times, say $Q$ \citep{Topology=000020of=000020gauge=000020fields=000020and=000020condensed=000020matter,Topology=000020from=000020the=000020differentiable=000020viewpoint,Unification=000020of=000020topological=000020invariants=000020in=000020Dirac=000020models}.
Suppose that $\hat{\mathbf{n}}$ has defects at $N$ different points
of $P$, where these defects possess the topological numbers $k_{1},\ldots,k_{N}$.
Then the Riemann-Hurwitz formula states that
\begin{equation}
2p-2=\left(2s-2\right)Q+\sum_{i=1}^{N}\left(k_{i}-1\right),\label{eq:RH}
\end{equation}
where $p=\mathrm{genus}\left(P\right)$ and $s=\mathrm{genus}\left(S\right)$
\citep{Gesammelte=000020mathematische=000020Werke=000020und=000020wissenschaftlicher=000020Nachlass,Ueber=000020Riemann'sche=000020Fl=0000E4chen=000020mit=000020gegebenen=000020Verzweigungspunkte,Iteration=000020of=000020rational=000020functions=000020Complex=000020analytic=000020dynamical=000020systems,Riemann=000020surfaces}.

In our case, $\hat{\mathbf{n}}\left(\mathbf{r}\right)$ is a mapping
from the unit cell $P$ on the plane to the sphere of spin states
$S$. The genus of $P$ is $p=1$ as the unit cell is topologically
equivalent to a torus, and the genus of $S$ is $s=0$. Therefore,
the Riemann-Hurwitz formula \eqref{eq:RH} yields
\begin{equation}
2Q=\sum_{i=1}^{N}\left(k_{i}-1\right).\label{eq:number}
\end{equation}
This formula leads to a significant conclusion: stable defects with
$k\geq2$ \emph{must} exist in the system. The formula further emphasizes
that a $k=1$ texture does not describe a defect; $k=1$ points do
not contribute to the Riemann-Hurwitz formula. It is important to
note that these defects appear not only in excited states, but are
intrinsic features manifesting even in the ground state of rotating
systems. This parallels the presence of point defects in the ground
state of an ordinary rotating superfluid.

The Riemann-Hurwitz formula provides the number of defects $N$ in
each unit cell. Specifically, if all the defects share the same $k$
value, the number of defects in each unit cell is
\begin{equation}
N=\frac{2Q}{k-1}.\label{eq:rh}
\end{equation}
Using Eq. \eqref{eq:intMH}, we find the density of defects in the
system:
\begin{equation}
\frac{N}{A}=\frac{1}{\pi F\left(k-1\right)}\frac{m\omega}{\hbar}.\label{eq:dens}
\end{equation}
Note that $Q$ does not appear in this identity. For a specific condensate
rotating at angular velocity $\omega$, the density of the realized
defects depends only on their $k$ value.

The formula for the density of defects remains the same for finite
$F$ values, since the number of defects per unit cell is unchanged
(the Riemann-Hurwitz formula applies even if the Jacobian of the mapping
is not constant, as long as its sign is) and the integration result
of the Mermin-Ho relation remains valid \citep{future}. Notably,
for $F=1$ and $k=2$, this formula coincides with the Feynman relation
for ordinary vortex lattices \citep{Application=000020of=000020quantum=000020mechanics=000020to=000020liquid=000020helium}.
We would like to again underscore the significance of the results
derived from the remarkable Riemann-Hurwitz formula: it provides us
with a prediction of an entirely new class of topological defects,
distinct from conventional topological point defects characterized
by the fundamental homotopy group.

As previously discussed, describing the spin texture for $F=\infty$
involves finding an equal-area mapping from the unit cell to the sphere.
Although infinitely many such mappings exist, explicitly constructing
one remains a challenging task. This paper focuses on explaining the
topological structure of the mappings and how considering the topological
defects aids in understanding it.

We assume $Q$ should be as small as possible for the ground state,
in order to achieve a simple topology. For a constant-sign Jacobian,
every point in $S$ has $Q$ preimages in $P$. Thus, it is not possible
to have $Q=1$, as it would describe a $1$-to-$1$ mapping, a homeomorphism,
which contradicts the topological inequivalence between the torus
and the sphere.

In order to find a $Q=2$ mapping, we examine the arrangement of the
point defects, since their positions and their images on the sphere
determine qualitatively the rest of the mapping. Assuming all the
defects of the mapping are of the simplest kind, namely $k=2$, the
Riemann-Hurwitz formula \eqref{eq:rh} implies that each unit cell
must contain $N=4$ defects. The mapping can take various forms, each
associated with different energy dictated by the second term in Eq.
\eqref{eq:energy} (the first term vanishes for area-preserving mappings).
To identify the ground state mapping and find the defect positions,
one must minimize this energy term. However, since ground states typically
exhibit higher symmetry, and considering the rotational symmetry inherent
in the problem, it is reasonable to focus on defect configurations
with greater symmetry. A natural choice is to arrange the defects
on the sphere in a tetrahedral configuration.
\begin{figure*}[t]
\centering
\begin{centering}
\includegraphics[scale=0.625]{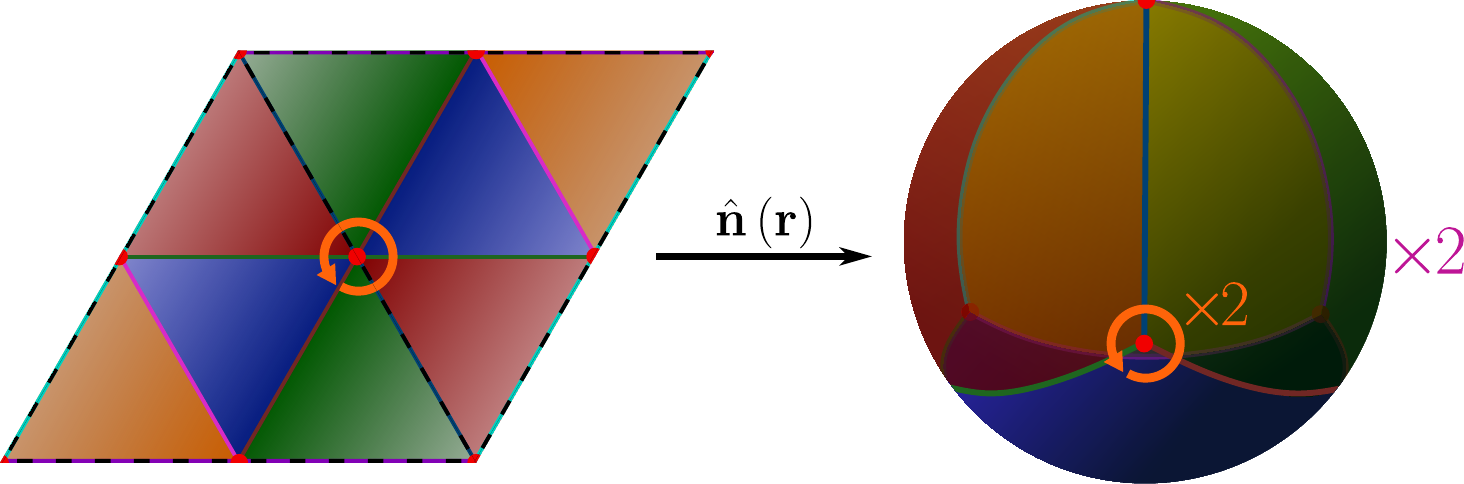}
\par\end{centering}
\caption{The magnetization texture mapping $\hat{\mathbf{n}}\left(\mathbf{r}\right)$
is obtained by dividing the plane into triangles and mapping them
to spherical triangles indicated by matching colors (the orange triangle
is in the back of the sphere). $k=2$ defects (red points) are located
at the vertices of the triangles, with $N=4$ of them per unit cell.
The images of the defects are arranged in the shape of a tetrahedron
on the sphere. Each unit cell covers the sphere $Q=2$ times, as the
two triangles outlined in dashed black lines each close up to form
one tetrahedron.\protect\label{fig:Four-sections-on}}
\end{figure*}

Now that the four defects are in place, we divide the sphere into
four spherical triangles with the defect points at their corners (see
Fig. \eqref{fig:Four-sections-on}). This configuration is then straightened
out into a tetrahedron with flat faces in some equal-area manner.
The four faces of the tetrahedron are then unfolded to form a triangle
in the plane (dashed black lines in Fig. \eqref{fig:Four-sections-on}).
Placing two such triangles side by side with reversed orientations
forms a parallelogram unit cell, which can be repeated to tile the
entire plane. This construction defines a mapping from the sphere
to the plane, which can be inverted to obtain $\hat{\mathbf{n}}\left(\mathbf{r}\right)$.
The resulting map is double covering ($Q=2$) because each parallelogram
unit cell consists of the two triangles, with each triangle covering
the sphere once under the mapping. On the plane, the defects are the
points around which the corresponding faces of the tetrahedron are
repeated twice, resulting in a $2$-to-$1$ texture around them.

Numerical minimization of energy \eqref{eq:energy} using a steepest
descent algorithm confirms that the actual ground state has the same
topological structure as the mapping shown in Fig. \eqref{fig:Four-sections-on},
manifesting a triangular lattice with tetrahedrally-arranged defects
on the sphere. This structure remains the same for all spin values.
These results are consistent with the prediction of a triangular lattice
for pseudospin-$1/2$ and spin-$1$ systems in Refs. \citep{Two-component=000020Bose-Einstein=000020condensates=000020with=000020a=000020large=000020number=000020of=000020vortices,Rotating=000020spin-1=000020bosons=000020in=000020the=000020lowest=000020Landau=000020level},
and its experimental observation for pseudospin-$1/2$ \citep{Vortex-lattice=000020dynamics=000020in=000020rotating=000020spinor=000020Bose-Einstein=000020condensates}.
However, the previous analyses have predominantly focused on separate
spinor components, unlike our approach, which adopts a geometrical
$\mathrm{SO}\left(3\right)$ symmetric viewpoint and reveals the presence
and significance of the defects.

Thus far, our analysis has focused on understanding the defects through
their influence on the magnetic structure. However, due to the Mermin-Ho
relation, they also have a crucial effect on the condensate flow.
To explore this, we relax the constraint $\mathbf{v}=\mathbf{\boldsymbol{\omega}}\times\mathbf{r}$,
allowing us to explore the ground state of a system with large yet
finite $F$, utilizing the Euler-Lagrange equations derived from energy
\eqref{eq:energy}.

It can be shown that for an area-preserving mapping the second spatial
derivatives of $\hat{\mathbf{n}}$ diverge at the defects, causing
the torque $\delta E/\delta\phi$ acting on $\hat{\mathbf{n}}$ to
diverge. Consequently, such a mapping is valid only for infinite $F$;
for any finite $F$, a different solution is required near the defects.
To address this, we study the cores of the defects, defined as the
regions where the area-preserving approximation breaks down significantly,
and characterized by a notable deviation of the vorticity from the
rigid rotation value $2\boldsymbol{\omega}$, as can be seen in Fig.
\eqref{fig:The--lattice.}. This definition differs from the definition
of a typical defect core, such as in scalar superfluid vortices, which
is based on a significant reduction in density. In this sense, the
defects we describe are coreless.

In the large-$F$ limit, the core size is found to be independent
of $F$. Since areas are scaled by $1/F$ in the mapping to the unit
sphere according to Eq. \eqref{eq:intMH}, the area corresponding
to the image of the core on the sphere is proportional to $F^{-1}$,
hence its angular size scales as $\phi\sim F^{-\frac{1}{2}}$ (when
$\hat{\mathbf{n}}$ at the defect is rotated to the north pole). For
large $F$, this region is small and thus approximately flat, allowing
us to simplify the Euler-Lagrange equations in this region by neglecting
the curvature of the sphere. Assuming a rotationally invariant structure
in the vicinity of the defect, we may take $\chi=k\alpha$ in order
to describe a $k$-defect. The approximated Euler-Lagrange equation
for $\phi$ is then:

\begin{equation}
x\frac{\partial}{\partial x}\left(x\frac{\partial u}{\partial x}\right)=k^{2}\left[\left(1-x^{2}\right)u+u^{3}\right],\label{eq:key2}
\end{equation}
where $u=\sqrt{F}\phi$, and $\mathbf{x}=\sqrt{m\omega/\hbar k}\mathbf{r}$
is a dimensionless form of the coordinates. Since this equation is
independent of the parameters, the typical scales for $u$ and $x$
are of order $1$, justifying the scales mentioned earlier. In the
original variables, the core area on the plane is of order $m\omega/\hbar$;
therefore, for large $F$, the unit cell size (Eq. \eqref{eq:intMH})
is much larger than the core size. Hence, the cores are far apart,
validating the analysis of each core separately. While $u\propto x$
near the defect for the area-preserving case, the solution of Eq.
\eqref{eq:key2} yields $u\propto x^{k}$. This corrects the aforementioned
singularities in the second derivatives of $\hat{\mathbf{n}}$, resulting
in an infinitely differentiable magnetization texture.

\begin{figure}[h]
\centering
\begin{centering}
\includegraphics[scale=0.45]{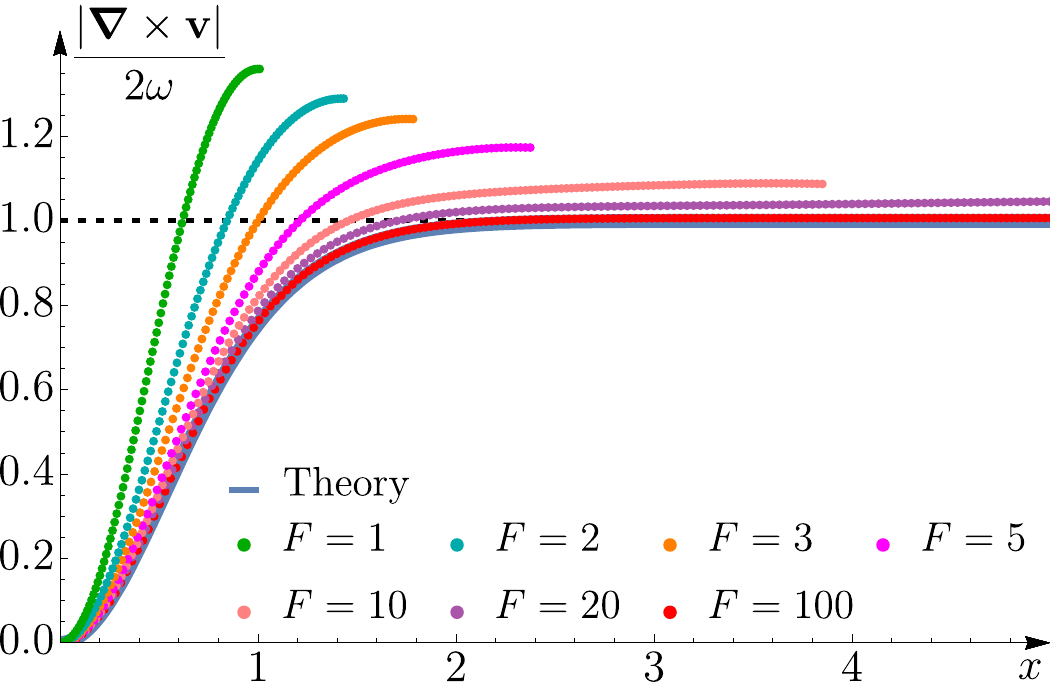}
\par\end{centering}
\caption{Blue curve: vorticity around a $k=2$ defect, calculated from Eq.
\eqref{eq:=000020vor} using the numerical solution of \eqref{eq:key2}.
Points: numerical gradient descent simulation results for the angle-averaged
vorticity around a $k=2$ defect for various values of $F$.\protect\label{fig:The-vorticity-gotten}}
\end{figure}

After rescaling, the Mermin-Ho relation around the defect becomes
\begin{equation}
\boldsymbol{\nabla}\times\mathbf{v}=2\boldsymbol{\omega}\frac{u}{x}\frac{\partial u}{\partial x}.\label{eq:=000020vor}
\end{equation}
Numerical solution of Eq. \eqref{eq:key2} for $u\left(x\right)$
yields the vorticity inside the core, shown in Fig. \eqref{fig:The-vorticity-gotten}.
The vorticity grows from $0$ to $2\omega$ on the scale of $x\approx1$,
consistent with the expected core size. Comparison with vorticity
of the defects in the numerical results shows convergence towards
the predicted curve as $F$ increases. Besides leading to a finite
core size, a finite $F$ has interesting effects on the vorticity
outside the core: its mean exceeds $2\omega$ (see Fig. \eqref{fig:The-vorticity-gotten})
and it is smaller along valleys connecting the defects (see Fig. \eqref{fig:The--lattice.}).
These corrections will be addressed in \citep{future}.

The distinction between the velocity field defects in our system and
conventional superfluid vortices is evident. Here, vorticity increases
gradually from zero at the defect point to $2\boldsymbol{\omega}$
with distance, whereas in regular superfluid vortices all the vorticity
is concentrated within the defect itself. This analysis, supported
by independent numerical results, shows that the positions of these
``unvortices'' align precisely with the locations of the magnetic
texture defects. This alignment establishes a direct connection between
the defects of $\mathbf{v}\left(\mathbf{r}\right)$ and $\hat{\mathbf{n}}\left(\mathbf{r}\right)$,
despite their fundamentally different nature.\\

We would like to thank Hillel Aharoni for suggesting the term ``unvortex'',
and Sandro Stringari for promoting initial interest in the problem
of rigidly rotating superfluids. RB acknowledges support from the
European Union\textquoteright s Seventh Framework Programme under
Grant No. PCIG-GA2013-631002. AMT acknowledges support from the Israeli
Science Foundation (ISF) Grant No. 1939/18. RB and AMT thank the hospitality
of the Aspen Center for Physics under Grant No. PHYS-1066293.


\begin{thebibliography}{99}
\bibitem[foot, 1]{foot-1}The contributions of centrifugal forces
and trap potential may disrupt the assumption of constant density
profile. However, except near the edges of the condensate, these effects
can be neglected due to strong interatomic interactions, rendering
them absent from the energy functional.
\end{thebibliography}
\end{document}